\documentclass{article}

\usepackage{fullpage}
\usepackage{parskip}
\usepackage{setspace}
\usepackage{mathtools}
\usepackage{tikz}
\usetikzlibrary{arrows}
\usetikzlibrary{decorations.markings}
\usetikzlibrary{calc}
\usetikzlibrary{shapes}
\usepackage{standalone}
\usepackage{float}
\usepackage{caption}
\usepackage{subcaption}
\usepackage{amsmath}
\usepackage{amsfonts}
\usepackage{amsthm}
\usepackage[ruled]{algorithm2e}
\usepackage{adjustbox}
\usepackage{enumerate}
\usepackage{hyperref}
\usepackage[nocompress]{cite}
\usepackage{authblk}
\usepackage[cachedir=_minted-ciw_paper, finalizecache=false, frozencache=true]{minted}
\usepackage{booktabs}
\usepackage{multicol}
\usepackage{algorithm2e}
\usepackage{pifont}

\newcommand{\cmark}{\ding{51}}%
\newcommand{\xmark}{\ding{55}}%

\definecolor{bg}{RGB}{249, 251, 223}

\title{Ciw: An open source discrete event simulation library.}
\author{Geraint I. Palmer\thanks{Corresponding author Geraint Palmer, palmergi1@cardiff.ac.uk}}
\author{Vincent A. Knight}
\author{Paul R. Harper}
\author{Asyl L. Hawa}
\affil{\small{\textit{School of Mathematics, Cardiff University, Senghennydd Road, Cardiff, CF24 4AG}}}
\date{\today}

\begin{document}

\maketitle

\begin{abstract}
This paper introduces Ciw, an open source library for conducting discrete event
simulations that has been developed in Python.
The strengths of the library are illustrated in terms of best practice and
reproducibility for computational research.
An analysis of Ciw's performance and comparison to several alternative discrete
event simulation frameworks is presented.

\textit{Keywords:} reproducibility, discrete event simulation, open source,
python
\end{abstract}

At the time of submission, this article was being considered for publication in
the Journal of Simulation.

\section{Introduction}

The analysis of queueing systems, especially those arranged into networks, is
a standard approach to studying a variety of real life operational systems.
Discrete event simulation (DES) is an extremely popular and rapidly growing
method of analysing networks of queues
\cite{brailsfordetal09, gunalpidd10, robinson05}.

Reproducibility and replicability, described as ``the cornerstone of cumulative
science''~\cite{sandveetak13}, is critical in order to assert correct results
and build on the work of others~\cite{sandveetak13, hongetal15}.
In computational research this can be achieved by following a number of best
practices~\cite{hongetal15, sandveetak13, prlicprocter12, cricketal14,
wilsonetal14, aberdour07, jimenezetal17, benureauetal17}.
A popular software paradigm for research is open source software which implies
software source code that is freely useable and modifiable.
In a recent review of open source discrete event simulation
software~\cite{dagkakisheavey16}, 44 open source discrete event simulation
solutions were found and reviewed, however not all followed best practice: 14
were found to have no available documentation, and over half failed to use any
version control.
The paper did not consider automated testing of these packages.

This paper introduces the open source Python library Ciw, which aims to enable
best practices within the domain of discrete event simulation, and yield
reproducible results.
Ciw is a Python library for the simulation of open queueing networks.
The core features of this library include the capability to simulate networks of
queues~\cite{jackson57}, multiple customer classes~\cite{kelly75}, and
restricted networks exhibiting blocking~\cite{onvuralperros86}.
A number of other features are also implemented, including priorities
\cite{cobham54}, baulking~\cite{anckerjrgafarian63a, anckerjrgafarian63b},
schedules~\cite{doshi86}, and deadlock detection \cite{palmeretal}.

This paper is structured as follows:
Section~\ref{sec:motivation} will provide full motivation for the library's use
and development, and Section~\ref{sec:features} will outline the features
currently implemented in the package.
Then, in Section~\ref{sec:architecture}, we will briefly discuss the code's
object-orientated structure and event-scheduling simulation algorithm, which
will be followed by an example of Ciw's usage and syntax in
Section~\ref{sec:example}.
In Section~\ref{sec:uses}, we will list how the library has been used in
academic work to date, and finally Section~\ref{sec:comparisons} will compare
Ciw with five other simulation frameworks.

\section{Motivation}\label{sec:motivation}

\begin{figure}
    \begin{center}
        \begin{tikzpicture}

\draw[ultra thick, -triangle 90] (0, 0) -- (20, 0);

\node[align=center] (pl) at (4, -1) {Programming\\Languages};
\node[align=center] (nogui) at (10, -2) {Simulation Packages\\without GUI};
\node[align=center] (gui) at (16, -1) {Simulation Packages\\with GUI};

\draw[->] (pl) -- (0, -1);
\draw[->] (pl) -- (8, -1);

\draw[->] (nogui) -- (6, -2);
\draw[->] (nogui) -- (14, -2);

\draw[->] (gui) -- (12, -1);
\draw[->] (gui) -- (20, -1);

\draw[-triangle 45] (2, 1.5) -- (2, 0);
\draw[-triangle 45] (6, 1.5) -- (6, 0);
\draw[-triangle 45] (9, 1.5) -- (9, 0);
\draw[-triangle 45] (11, 1.5) -- (11, 0);
\draw[-triangle 45] (14, 1.5) -- (14, 0);
\draw[-triangle 45] (18, 1.5) -- (18, 0);

\node[align=center] at (2, 2) {C++};
\node[align=center] at (6, 2) {Spreadsheet\\Modelling};
\node[align=center] at (9, 2) {SimPy};
\node[align=center] at (11, 2) {Ciw};
\node[align=center] at (14, 2) {AnyLogic};
\node[align=center] at (18, 2) {SIMUL8};

\end{tikzpicture}
    \end{center}
    \caption{A suggested spectrum corresponding to user interface,
        with illustrative positioning for six simulation options.}
    \label{fig:spectrum}
\end{figure}
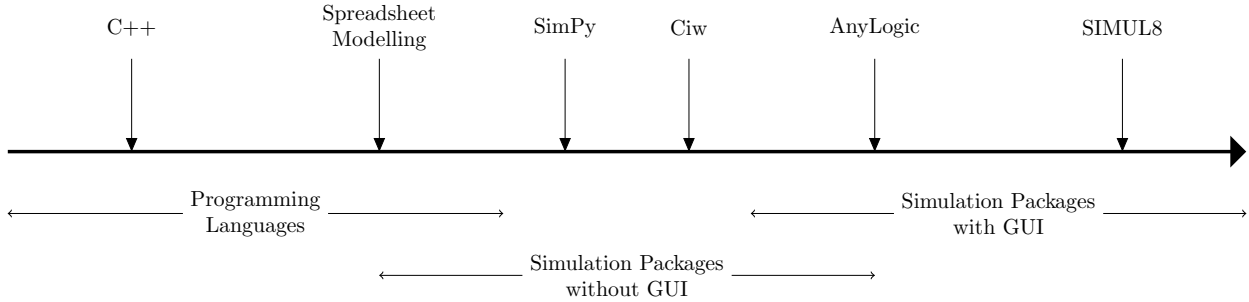

Simulation options traditionally fall into discrete categories
\cite{law07, robinson14}, consisting of programming languages, simulation
packages, and spreadsheet modelling.
We consider simulation frameworks on a spectrum corresponding to the user
interface.
Such a spectrum is shown in Figure~\ref{fig:spectrum}, with some suggested
positions for a selection of simulation options, including Ciw, SimPy
\cite{simpy}, AnyLogic \cite{anylogic}, SIMUL8 \cite{simul8}, as well as
building a simulation in C++, and spreadsheet modelling.

Advantages and disadvantages of these methods have been discussed extensively
\cite{law07, robinson14, dagkakisheavey16, bellokeefe87}.
Programming from scratch is considered more flexible, is bespoke, may improve
speed, and increases the variety of performance measures collected.
However, a lack of user interface may hinder model communication.
It is discussed in some literature that simulation packages, especially those
with a graphical user interface (GUI), are more accessible, easily modifiable,
and easier to communicate with non-specialists.
GUIs can aid with conceptual modelling, model validation and verification
\cite{bellokeefe87, kirkpatrickbell89, beltonelder94}.

However some simulation packages come with several disadvantages for example
high costs (licences, training, plug-ins and maintenance), they often lack
modularity, low model reusability, and lack of access to the source code can
impede understanding, customisation and flexibility.
Furthermore, it is suggested \cite{bellokeefe87} that the addition of a GUI can
lead to bad
simulation practice.
This includes: disregarding formal methodology such as statistical analysis in
favour of watching animations; introducing bias in the model building process by
building models that represent how a system should work instead of how they
actually work; and false model validation in which realistic graphics imply a
realistic model.

Two themes arise when discussing research and software development:
reproducibility and sustainability, and best practice \cite{hongetal15,
sandveetak13, prlicprocter12, cricketal14, wilsonetal14, aberdour07,
jimenezetal17, benureauetal17}.
The Ciw library aims to enable users to meet these standards in the domain
of discrete event simulation.

In~\cite{kilgore01} three properties are listed as minimum requirements to
ensure reproducibility in simulation:

\begin{itemize}
  \item Readability
  \item Modularity
  \item Extendibility
\end{itemize}

All three properties can apply to Python~\cite{python}, the ecosystem in which
modelling with Ciw takes place.
Python is an open source, object-orientated, high level language.
Python's advantages as a language for developing simulation models are listed in
\cite{dagkakisheavey16}.
These include its intuitive and readable syntax, and potential to form a
community of users and developers.
In addition, Python is attractive to researchers due to the extensive collection
of other scientific libraries available to integrate work, for example combining
simulation with data analysis, machine learning, and optimisation.
An example of this as a teaching exercise with Ciw is seen in \cite{knight17}.
There are a number of popular scientific Python libraries, including:

\begin{itemize}
  \item NumPy~\cite{numpy} and SciPy~\cite{scipy01} for scientific computation
  \item scikit-learn~\cite{scikitlearn} for machine learning and optimisation
        algorithms
  \item Pandas~\cite{pandas} for data analysis tools
  \item Matplotlib~\cite{matplotlib} and Seaborn~\cite{seaborn} for data
        visualisation
  \item SymPy~\cite{sympy} for symbolic mathematics
\end{itemize}

In~\cite{wilsonetal14} the authors discuss the advantages of using high level
programming languages such as Python for research software over low level
languages like C and Fortran.
Advantages include an increase in productivity when writing in high level
languages, better readability, and rapid design decisions and prototyping.
A downside however, due to the fact that Python is an interpreted language, is
that the computational speed will not be as fast as compiled languages.

Another strength of Ciw is that it is completely open source.
It has one of the most flexible and permissive licences, the MIT licence, and
is written in an open source language.
This offers the user immediate advantages over commercial-off-the-shelf
simulation packages.
All source code is available for inspection, testing, and modification.
This enables and encourages greater understanding of the underlying methodology,
increases model confidence, and provides an extendible framework in which
discrete event simulation may be carried out.
Furthermore, the elimination of license fees and maintenance costs facilitate
model sharing,  open science, and reproducibility.
This overcomes common problems with commercial software with more stringent
licences, where models may sometimes not be shared between two computers with
the same software.

This ecosystem provides an opportune way in which reproducible scientific
research can be conducted:

\begin{itemize}
  \item All manual data manipulation can be avoided~\cite{sandveetak13,
  cricketal14, wilsonetal14}.
  \item All raw data can be saved~\cite{sandveetak13}.
  \item All models can be version controlled~\cite{sandveetak13, wilsonetal14,
  benureauetal17}.
  \item All models can be scrutinised by automated testing~\cite{wilsonetal14,
  benureauetal17}.
  \item All models can be shared~\cite{cricketal14, hongetal15, sandveetak13,
  jimenezetal17, benureauetal17}.
\end{itemize}

Ciw is also developed in a sustainable manner, and strives to follow best
practice in research software development.
This includes extensive testing (it has 100\% test coverage~\cite{coverage}),
comprehensive documentation, readability, modularisation, transparency, and use
of version control~\cite{wilsonetal14, prlicprocter12}.

Object-orientation, an important feature of Python, lends itself well to
simulation~\cite{law07, dagkakisheavey16}.
In~\cite{dagkakisheavey16} the authors state that ``DES is a traditional
paradigm where object orientation is intuitively adopted''.
The argument for linking object-orientation to one particular method of discrete
event simulation, the three-phase approach, is given in~\cite{pidd95}.
Breaking a simulation down into events, activities and entities, as is required
for the three-phase approach, is a form of modularisation itself.
This equally applies to the similar event scheduling approach used in Ciw.
In addition, simulation modellers habitually think of entities as belonging to a
class, or classes, of similar entities.
It is intuitive to build systems like this in an environment where
modularisation is key, such as in an object-orientated programming language.
Further advantages of using object-orientation are listed in~\cite{law07}: its
flexibility, its ability to deal with complexity through modularisation, and its
high reusability.

As stated previously, using open source software provides distinct advantages
over traditional commercial-off-the-shelf simulation packages.
Similarly, open source development can provide many advantages over closed
source development.
However,~\cite{dagkakisheavey16} argues that apart from eliminating the licence
fee, simply being open source does not offer immediate advantages for
developers, but it is the ethos and culture that comes with open source that
provide the advantages.
It is argued in~\cite{vonkroghvonhippel06} that open source culture provides
incentives to innovate, as there is no need for a large demand or promise of
recoupment of financial investment for certain features to be developed.
That is, private needs create public goods.
This has been evident in Ciw, where new research can be directly implemented
into the software and tested and experimented quickly (see
Section~\ref{sec:uses}). New features have been implemented after discussions
with users from around the world via the online issue tracker.
Freedom of development is another crucial aspect of open source according to
\cite{vonkroghvonhippel06}, where users can fork and develop their own versions
of software for their bespoke needs.
An argument for promoting best practice in open source software is given in
\cite{aberdour07}, as it achieves better quality software.
Some of these best practices arise naturally in an open source environment, for
example rapid release cycles, code reviews, and code modularity.
Open source development actually encourages these best practices due to its
transparency and the opportunities for developers to showcase their work
\cite{jimenezetal17}.

A review of open source discrete event simulation software is given in
\cite{dagkakisheavey16}, and as mentioned previously, a number of frameworks
failed to follow best practice in their development.
Three Python libraries were found, though only one was found to meet the quality
requirements of the study, SimPy~\cite{simpy}.
Further,~\cite{kilgore01} draws many parallels between open source development
and simulation modelling, while concluding that the ``steady, long-term progress
toward libraries of easily extendible and easily reusable simulation code'' is
an important direction for simulation modellers.

To summarise, Ciw is an object-orientated, open source Python library with the
following qualities:
\begin{itemize}
  \item Open, accessible source code promotes understanding, development and
  modification. Online issue tracker and open development environment fosters
  discussion, idea generation, and development. Permissive licence allows it to
  be extended, modified and shaped to the users' needs.
  \item Code development follows best practice guidelines for reproducibility,
  code quality, and modification. Modularity allows modification and extension
  through inheritance.
  \item Python ecosystem allows it to be used flexibly within the programming
  language, allowing ease of experimentation and integration with other
  scientific tools. Models can be tested and version controlled.
  \item Models are readable and the package has extensive bilingual
  documentation, to enhance model communication.
\end{itemize}

\section{Features}\label{sec:features}

Ciw's main functionality is the simulation of open restricted queueing networks
that exhibit blocking, and supports multiple classes of customer:

\begin{itemize}
  \item A \textit{queueing network} is a system consisting of a number of
  service centres where customers may wait in a queue for service; connected by
  a transition matrix of probabilities $r_{ij}$, the probability of joining node
  $j$ after completing service at node $i$.
  \item A queueing network is described as \textit{open} if customers can leave
  the system, and new customers can arrive from outside the system
  \cite{stewart09}.
  \item A queueing network is described as \textit{restricted} if nodes have
  limited queueing capacity, that is, only room for a certain amount of
  customers to wait at any one time. If a node's queueing capacity is full, then
  external arrivals are rejected, and \textit{Type I blocking}
  \cite{onvuralperros86} occurs for customers transitioning from other nodes.
  That is, after service they remain with their server until space becomes
  available at their destination node, while that server is unavailable to serve
  any other customer.
  \item \textit{Multiple classes of customer} refers to the possibility of
  having more than one type of customer using and sharing the same resources,
  but using them in different ways. For example, each class of customer may have
  its own distinct inter-arrival time distributions, service time distributions,
  and transition matrices.
\end{itemize}

In addition to these main properties, Ciw can simulate a number of other
features:

\begin{itemize}
  \item \textit{A choice of inter-arrival and service time distributions}:
  Including Uniform, Deterministic, Triangular, Exponential, Gamma, Truncated
  Normal, Lognormal, Weibull, and the possibility of users defining their own
  Discrete, Continuous, Empirical, Sequential, and Time Dependant distributions.
  \item \textit{Batch arrivals:} At each external arrival, a number of customers
  may arrive simultaneously. Ciw allows sampling from a discrete probability
  distribution to obtain batch sizes.
  \item \textit{Priority classes:} A mapping from customer classes to priority
  levels. This allows customers with higher priority to jump ahead of customers
  with lower priority each time they enter a queue.
  \item \textit{Baulking customers:} At each external arrival, customers have a
  probability $b(m)$ of baulking (choosing not to join the system), given that
  there are $m$ customers already at that node. Ciw allows users to define their
  own baulking function $b(m)$ as a Python function.
  \item \textit{Server schedules:} Cyclic server schedules may be defined for
  each node, that is the number of servers at a node may increase or decrease as
  servers go on and off duty at fixed times during the simulation run.
  \item \textit{Dynamic customer classes:} After service at a node, customers
  may re-sample their customer class according the a class change matrix of
  probabilities $p_{ij}$, the probability of a customer of class $i$ becoming a
  customer of class $j$ after service at that node. This means that their
  behaviour (service distributions, transition matrices, and priorities) will
  also change.
  \item \textit{Deadlock detection:} Restricted queueing networks with cycles
  can cause the phenomenon of deadlock \cite{palmeretal}. Traditionally
  deadlocks are difficult to detect, however Ciw has the ability to terminate a
  simulation run once deadlock has been reached, using the state digraph method.
\end{itemize}

Ciw also offers a number of termination conditions:

\begin{itemize}
  \item Simulating until a maximum amount of time has passed.
  \item Simulating until a maximum number of customers have arrived/been
  accepted/finished.
  \item Simulating until deadlock has been reached.
\end{itemize}

\section{Architecture}\label{sec:architecture}

Ciw makes full use of Python's object-orientated nature:

\begin{itemize}
  \item A \texttt{Simulation} object is a one-use object used for one run of a
  simulation, which contains a network of \texttt{Node} objects.
  \item Each \texttt{Node} object contains \texttt{Server} objects.
  \item \texttt{Individual} objects are passed around the network of
  \texttt{Node} objects, where they wait to be served by \texttt{Server} objects.
  \item Each \texttt{Individual} object carries a number of named tuples that
  record the history of a single service.
  \item The \texttt{ArrivalNode} creates new \texttt{Individual} objects to
  enter the simulation.
  \item The \texttt{ExitNode} collects \texttt{Individual}s that leave the
  system.
  \item \texttt{Network} objects, that also consist of \texttt{ServiceCentre}
  and \texttt{CustomerClass} objects, define a queueing network in order to
  create a \texttt{Simulation} object.
\end{itemize}

\begin{figure}
    \begin{center}
        \begin{tikzpicture}

\node[align=center] (simulation) [style={minimum height=1.5cm, minimum width=3.5cm, draw=black,fill=lime!80!black,text=black,shape=ellipse,fill opacity=0.6, text opacity=1}] at (0, 0) {\LARGE Simulation};

\node[align=center] (nodes) [style={minimum height=1.5cm, minimum width=3.5cm, draw=black,fill=lime!80!black,text=black,shape=ellipse,fill opacity=0.6, text opacity=1}] at (7.5, 3.75) {\LARGE Nodes};

\node[align=center] (arrivalnode) [style={minimum height=1.5cm, minimum width=3.5cm, draw=black,fill=lime!80!black,text=black,shape=ellipse,fill opacity=0.6, text opacity=1}] at (7.5, 0) {\LARGE Arrival Node};

\node[align=center] (exitnode) [style={minimum height=1.5cm, minimum width=3.5cm, draw=black,fill=lime!80!black,text=black,shape=ellipse,fill opacity=0.6, text opacity=1}] at (7.5, -3.75) {\LARGE Exit Node};

\node[align=center] (servers) [style={minimum height=1.5cm, minimum width=3.5cm, draw=black,fill=lime!80!black,text=black,shape=ellipse,fill opacity=0.6, text opacity=1}] at (14, 1.875) {\LARGE Servers};

\node[align=center] (individuals) [style={minimum height=1.5cm, minimum width=3.5cm, draw=black,fill=lime!80!black,text=black,shape=ellipse,fill opacity=0.6, text opacity=1}] at (14, -1.875) {\LARGE Individuals};

\node[align=center] (network) [style={minimum height=1.5cm, minimum width=3.5cm, draw=black,fill=orange!50!yellow,text=black,shape=ellipse,fill opacity=0.6, text opacity=1}] at (-7.5, 1.875) {\LARGE Network};

\node[align=center] (servicecentre) [style={minimum height=1.5cm, minimum width=3.5cm, draw=black,fill=orange!50!yellow,text=black,shape=ellipse,fill opacity=0.6, text opacity=1}] at (-14, 3.75) {\LARGE Service Centre};

\node[align=center] (customerclass) [style={minimum height=1.5cm, minimum width=3.5cm, draw=black,fill=orange!50!yellow,text=black,shape=ellipse,fill opacity=0.6, text opacity=1}] at (-14, 0) {\LARGE Customer Class};

\node[align=center] (detector) [style={minimum height=1.5cm, minimum width=3.5cm, draw=black,fill=cyan!96!blue,text=black,shape=ellipse,fill opacity=0.6, text opacity=1}] at (-3.75, -3.5) {\LARGE Deadlock Detector};

\node[align=center] (tracker) [style={minimum height=1.5cm, minimum width=3.5cm, draw=black,fill=cyan!96!blue,text=black,shape=ellipse,fill opacity=0.6, text opacity=1}] at (-7.5, -1.875) {\LARGE State Tracker};

\draw (17.75, -1.5) rectangle (24, 2);
\draw[fill=lime!80!black, fill opacity=0.6] (18.75, 1) rectangle (20.2, 1.5);
\draw[fill=orange!50!yellow, fill opacity=0.6] (18.75, 0) rectangle (20.2, 0.5);
\draw[fill=cyan!96!blue, fill opacity=0.6] (18.75, -1) rectangle (20.2, -0.5);
\node[align=left] at (21, 1.25) {\Large Core};
\node[align=left] at (21.05, 0.25) {\Large Input};
\node[align=left] at (21.35, -0.75) {\Large Optional};

\draw (servers) edge[ultra thick,-triangle 90,out=-65,in=65] (individuals);
\draw (individuals) edge[ultra thick,-triangle 90,out=115,in=-115] (servers);
\draw (servers) edge[ultra thick,-triangle 90,out=135,in=0] (nodes);
\draw (individuals) edge[ultra thick,-triangle 90] (nodes);
\draw (individuals) edge[ultra thick,-triangle 90,out=180,in=-45] (arrivalnode);
\draw (individuals) edge[ultra thick,-triangle 90,out=-135,in=0] (exitnode);
\draw (nodes) edge[ultra thick,-triangle 90,out=-180,in=75] (simulation);
\draw (arrivalnode) edge[ultra thick,-triangle 90] (simulation);
\draw (exitnode) edge[ultra thick,-triangle 90,out=180,in=-75] (simulation);

\draw (network) edge[ultra thick,-triangle 90, out=0, in=150] (simulation);
\draw (servicecentre) edge[ultra thick,-triangle 90,out=0,in=135] (network);
\draw (customerclass) edge[ultra thick,-triangle 90,out=0,in=-135] (network);

\draw (tracker) edge[ultra thick,-triangle 90,out=45,in=180] (simulation);
\draw (detector) edge[ultra thick,-triangle 90] (simulation);

\draw[draw=gray] (customerclass) edge[ultra thick, dashed] (servicecentre);

\end{tikzpicture}
    \end{center}
    \caption{Ciw's architecture.}
    \label{ciw_code_structure}
\end{figure}
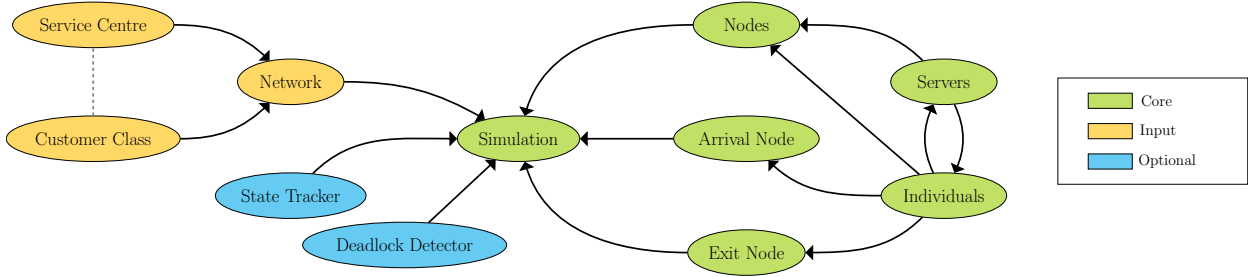

Figure~\ref{ciw_code_structure} summarises and categorises the interconnecting
objects that make up the Ciw framework.
Some optional objects, state trackers and deadlock detectors, are also used for
some features.

Ciw uses the event scheduling approach~\cite{robinson14}, similar to the popular
three-phase approach.
This deviates significantly from the other major Python alternative, SimPy, that
uses the process based approach.
In the event scheduling approach, three types of event take place:
A-Events move the clock forward, B-Events are pre-scheduled events, and C-Events
are events that arise because a B-Event has occurred.

Here A-Events correspond to moving the clock forward to the next B-Event.
B-Events correspond to either an external arrival, a customer finishing service,
or a server shift change.
C-Events correspond to a customer starting service, customer being released from
a node, and being blocked or unblocked.

In event-scheduling the following process occurs:

\begin{enumerate}
  \item Initialise the simulation.
  \item A-Phase: move the clock to the next scheduled event.\label{enum:2}
  \item Take a B-Event scheduled for now, carry out the event.\label{enum:3}
  \item Carry out all C-Events that arose due to the event carried out
  in \ref{enum:3}.\label{enum:4}
  \item Repeat \ref{enum:3} - \ref{enum:4} until all B-Event scheduled for that
  date have been carried out.\label{enum:5}
  \item Repeat \ref{enum:2} - \ref{enum:5} until a terminating criteria has been
  satisfied.
\end{enumerate}

Each \texttt{Node} object, including the \texttt{ArrivalNode}, has a
\texttt{have\_event} method and a \texttt{next\_event\_date} attribute.
The \texttt{next\_event\_date} attribute is updated each time an event occurs,
and corresponds to the date that the next B-Event at that object is scheduled to
happen.
At the A-Phase the simulation's clock is moved to the next
\texttt{next\_event\_date} of all the \texttt{Node} objects.

The \texttt{ArrivalNode}'s \texttt{have\_event} method spawns a new
\texttt{Individual}, and sends them to their appropriate \texttt{Node} (barring
baulking or exceeding node capacity).
All other \texttt{Node}s' \texttt{have\_event} method consist of one of two
events: an \texttt{Individual} finishing service at that node, or a shift change
for the \texttt{Server}s.

C-Events are not coded explicitly, but follow on naturally as consequences of
the B-Event described above.
For example when the \texttt{ArrivalNode} spawns a new \texttt{Individual} and
is successfully sent to a \texttt{Node}, if there is not another individual
waiting in the queue, that individual is attached to a \texttt{Server} and
begins service.
Similarly when an \texttt{Individual} finishes service, that individual is sent
to another \texttt{Node}, and may join a queue or begin service is a server is
free, while a waiting customer now begins service with the freed \texttt{Server}
at the current \texttt{Node}.

\section{Illustrative Use}\label{sec:example}

The library is installable from the Python package index, which means it is
readily available to anyone with Python \cite{python} (versions 2.7, 3.4 and
above) and an internet
connection.
The source code is available on GitHub: \url{https://github.com/CiwPython/Ciw},
under the MIT licence.
Full documentation is available (both in English and Welsh, in pdf and html
form), and hosted on Read The Docs: \url{http://ciw.readthedocs.io/}.

In order to demonstrate usage in more detail, consider the following system:

\begin{itemize}
  \item Two classes of jobs enter a computer repair clinic: scheduled jobs (S)
  and unscheduled jobs (U).
  \item Scheduled jobs arrive in batches of two, once per hour.
  \item Unscheduled jobs arrive randomly according to a Poisson distribution, at
  rate one per hour.
  \item Unscheduled jobs take priority over scheduled jobs, and thus join the
  queue ahead of scheduled jobs.
  \item The repair clinic consists of two nodes: an inspection desk with two
  servers where jobs arrive, and a repair room with one server.
  \item There is infinite queueing capacity at the inspection desk, but no
  queueing capacity between the inspection desk and the repair room, thus Type I
  blocking occurs here.
  \item All service times follow Exponential distributions: scheduled jobs take
  an average of 20 minutes to be inspected and 60 minutes to repair,
  unscheduled jobs take an average of 40 minutes to be inspected and 90 minutes
  to repair.
  \item 5\% of all scheduled jobs require repair and 40\% of all unscheduled
  jobs require repair.
\end{itemize}

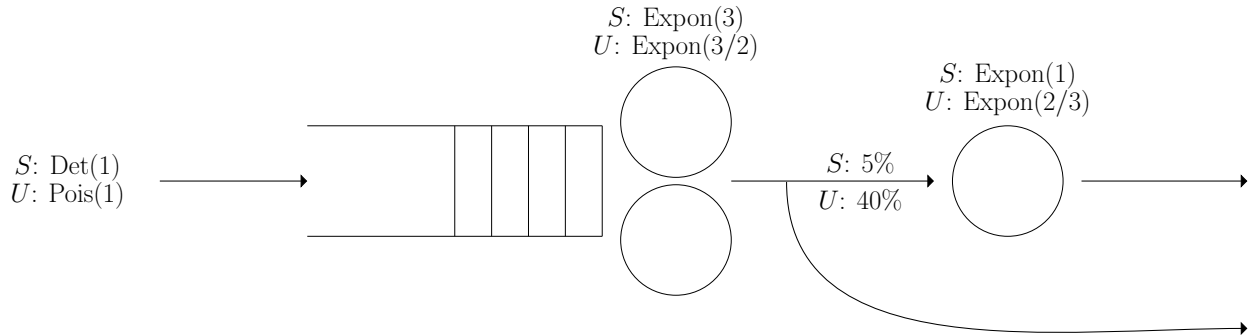
\begin{figure}
\begin{center}
\begin{tikzpicture}

  \node (exitstart) at (13, 1.6) {};

  \draw (0, 0) -- (8, 0);
  \draw (8, 0) -- (8, 3);
  \draw (0, 3) -- (8, 3);
  \draw (4, 0) -- (4, 3);
  \draw (5, 0) -- (5, 3);
  \draw (6, 0) -- (6, 3);
  \draw (7, 0) -- (7, 3);

  \draw (10, 3.1) circle (1.5);
  \draw (10, -0.1) circle (1.5);
  \node[align=center] at (10, 5.5) {\huge{$S$: Expon$(3)$}\\\huge{$U$: Expon$(3/2)$}};

  \draw (19, 1.5) circle (1.5);
  \node[align=center] at (19, 4) {\huge{$S$: Expon$(1)$}\\\huge{$U$: Expon$(2/3)$}};

  \draw[thick, -triangle 90] (11.5, 1.5) -- (17, 1.5);
  \draw[thick, -triangle 90] (21, 1.5) -- (25.5, 1.5);
  \draw[thick, -triangle 90] (-4, 1.5) -- (0, 1.5);
  \node[align=center] at (-6.5, 1.5) {\huge{$S$: Det$(1)$}\\\huge{$U$: Pois$(1)$}};
  \draw[thick, -triangle 90] (exitstart) edge[out=-90, in=180] (25.5, -2.5);

  \node[align=center] at (15, 1.5) {\huge{$S$: $5\%$}\\\\\huge{$U$: $40\%$}};

\end{tikzpicture}
\end{center}
\caption{Diagrammatic representation of the repair clinic example.
$S$ denotes scheduled jobs, $U$ denotes unscheduled jobs.}
\label{fig:repairclinic}
\end{figure}

The system is shown in Figure~\ref{fig:repairclinic}.
The repair clinic runs for 24 hours a day.
The example below will run a simulation of this system using Ciw and will obtain
estimates for the values of:

\begin{itemize}
  \item The average waiting time of unscheduled jobs at the inspection desk.
  \item The average time a job is spent blocked at the inspection desk
  (regardless of job class).
\end{itemize}

The code shown in Figure~\ref{fig:example_network} gives the code needed to
create the \texttt{Network} object that defines the system above.
The code in Figure~\ref{fig:example_run} runs the simulation over 20 trials, for
7 days, with a warm-up time of 1 day.

\begin{figure}
\begin{minted}[bgcolor=bg]{python}
>>> import ciw
>>> assert ciw.__version__ == '1.1.3'

>>> N = ciw.create_network(
...     Arrival_distributions={
...         'Class 0': [['Deterministic', 1.0],
...                     'NoArrivals'],
...         'Class 1': [['Exponential', 1.0],
...                      'NoArrivals']},
...     Service_distributions={
...         'Class 0': [['Exponential', 3.0],
...                     ['Exponential', 1.0]],
...         'Class 1': [['Exponential', 3.0/2.0],
...                     ['Exponential', 2.0/3.0]]},
...     Transition_matrices={
...         'Class 0': [[0.0, 0.05],
...                     [0.0, 0.0]],
...         'Class 1': [[0.0, 0.4],
...                     [0.0, 0.0]]},
...     Batching_distributions={
...         'Class 0': [['Deterministic', 2],
...                     ['Deterministic', 1]],
...         'Class 1': [['Deterministic', 1],
...                     ['Deterministic', 1]]},
...     Priority_classes={
...         'Class 0': 1,
...         'Class 1': 0},
...     Queue_capacities=['Inf', 0],
...     Number_of_servers=[2, 1]
... )

\end{minted}
\caption{Ciw code required to create the \texttt{Network} object for the
example system.}
\label{fig:example_network}
\end{figure}

\begin{figure}
\begin{minted}[bgcolor=bg]{python}
>>> def run_trial(seed, N):
...     ciw.seed(seed)
...     Q = ciw.Simulation(N)
...     Q.simulate_until_max_time(24*8)
...     recs = Q.get_all_records()
...     waits_unscheduled = [r.waiting_time for r in recs
...                          if r.customer_class==1
...                          if r.node==1
...                          if r.arrival_date > 24]
...     servers_blocked = [r.time_blocked for r in recs
...                        if r.node==1
...                        if r.arrival_date > 24]
...     average_wait = sum(waits_unscheduled) / len(waits_unscheduled)
...     average_blocked = sum(servers_blocked) / len(servers_blocked)
...     return average_wait, average_blocked

>>> waits = []
>>> blocked = []
>>> for seed in range(20):
...     average_wait, average_blocked = run_trial(seed, N)
...     waits.append(average_wait)
...     blocked.append(average_blocked)

>>> print(sum(waits) / len(waits))
2.08127848232403

>>> print(sum(blocked) / len(blocked))
0.23280830359597315

\end{minted}
\caption{Ciw code used to run the example system over 20 trials, and obtain
the average waiting time for unscheduled jobs, and the average time blocked
for all jobs.}
\label{fig:example_run}
\end{figure}

\section{Use cases}\label{sec:uses}

To date, Ciw has been used for various theoretical, practical and pedagogic
applications, including:

\begin{itemize}
  \item Theoretical work investigating deadlock in open restricted queueing
  networks by Palmer, G.I., Harper, P.R. and Knight, V.A. \cite{palmeretal}. A
  graph theoretical technique to detect deadlock during a run of a discrete
  event simulation was developed, and incorporated into the Ciw framework:
  \url{http://ciw.readthedocs.io/en/latest/Guides/deadlock.html}. Experiments on
  the time to reach deadlock were undertaken.

  \item The modelling of an ophthalmology clinic at Cardiff and Value University
  Health Board was undertaken by Morgan, J. and H\"{o}lscher, L., in order to
  investigate the best patient scheduling strategy. This project was essential
  to the development of Ciw, as many features were added to the library due to
  the requirements of the project: server schedules, dynamic customer classes,
  and exact arithmetic.

  \item Models of cancer patient diagnoses were built by Harper, P.R. and
  Palmer, G.I. for the Wales Cancer Network and Cwm Taf University Health Board.
  These models and what-if scenarios were used to advice national policy on
  capacity increases for diagnostic tests in Wales in order to reach potential
  Welsh government set cancer diagnosis time targets.

  \item A Nuffield research placement \cite{nuffieldresearchplacements} student
  Huang, C. undertook research with Knight V.A. and Palmer, G.I. studying
  deadlock in queueing networks, extending previous experiments to include
  baulking customers and servers taking vacations.

  \item The library has been used as part of a 2 day `hackathon' as part of the
  MSc in Operational Research and Applied Statistics at Cardiff University. The
  hackathon aims to increase familiarity with object-orientated programming by
  working on a Python project. In 2017, the project was to write a genetic
  algorithm to best configure three queues in series, using Ciw to obtain the
  cost function. An example solution is given in \cite{knight17}.
\end{itemize}

\section{Comparison with Other Simulation Frameworks}\label{sec:comparisons}

Five other popular simulation frameworks from across the spectrum corresponding
to user interface (see Figure~\ref{fig:spectrum}) were chosen for comparison.
They are compared on their appropriateness for conducting reproducible research
in the domain of discrete event simulation.
The frameworks chosen were:

\begin{multicols}{3}
\begin{itemize}
  \item C++ (version 11, compiled using g++ 4.2.1)
  \item Spreadsheet modelling (implemented in Microsoft Excel 2013)
  \item SimPy (version 3.0.10, using Python 3.5.1)
  \item Ciw (version 1.1.3, using Python 3.5.1)
  \item AnyLogic (AnyLogic 8 University 8.1.0)
  \item SIMUL8 (SIMUL8 2014 [Exclusive EDUCATIONAL SITE Edition])
\end{itemize}
\end{multicols}

A model of an $M/M/3$ queue~\cite{stewart09}, with arrival rate $\lambda = 10$,
and service rate $\mu = 4$ was built in each framework.
The models were run for $800$ time units, with a warmup time of $100$ time
units.
The aim was to find the average waiting time in the queue.

\begin{figure}
    \begin{center}
        \begin{subfigure}[b]{0.42\textwidth}
            \includegraphics[width=\textwidth]{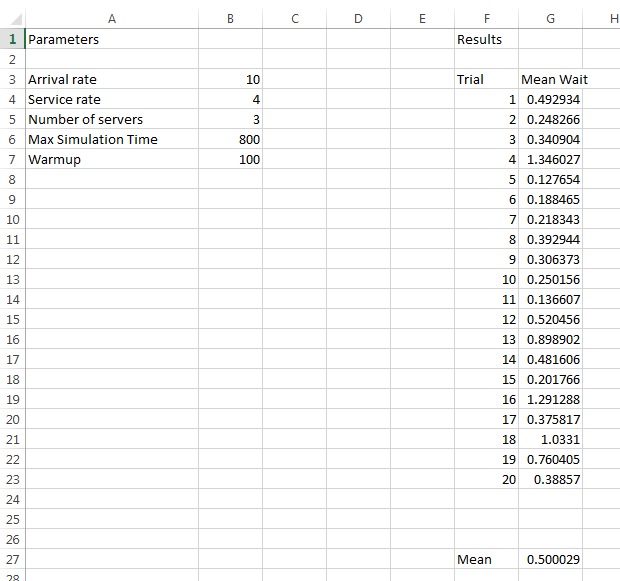}
            \caption{\label{fig:excelparams}}
        \end{subfigure}
        \hspace{0.1\textwidth}
        \begin{subfigure}[b]{0.42\textwidth}
            \includegraphics[width=\textwidth]{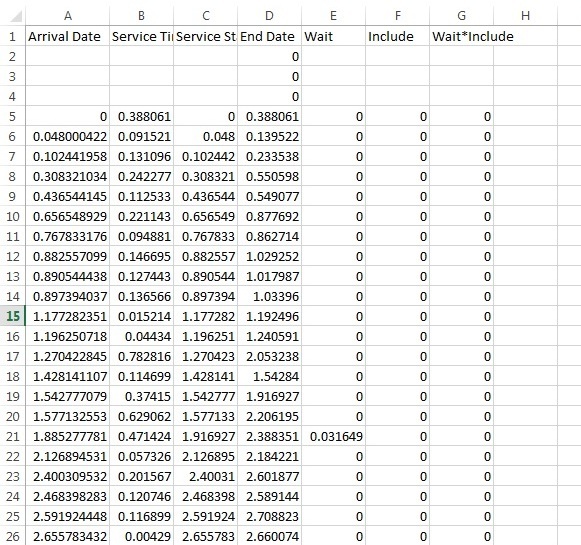}
            \caption{\label{fig:excelmechanics}}
        \end{subfigure}
    \end{center}
    \caption{Screenshots of the spreadsheet model developed in Microsoft Excel:
        Figure~\ref{fig:excelparams} shows parameters and results,
    Figure~\ref{fig:excelmechanics} shows the mechanics of one trial.}
    \label{fig:spreadsheetscreenshot}
\end{figure}

\begin{figure}
    \begin{center}
        \includegraphics[width=0.7\textwidth]{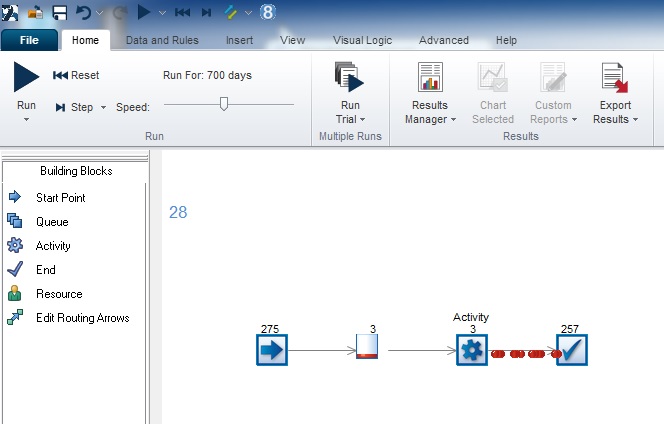}
    \end{center}
    \caption{Screenshot of the SIMUL8 model.}
    \label{fig:simul8screenshot}
\end{figure}

\begin{figure}
    \begin{center}
        \includegraphics[width=0.7\textwidth]{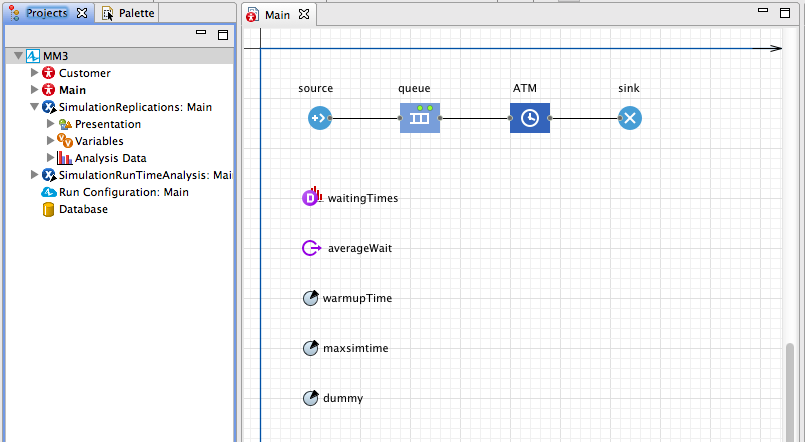}
    \end{center}
    \caption{Screenshot of the AnyLogic model.}
    \label{fig:anylogicscreenshot}
\end{figure}

The C++, SimPy, and Ciw models can be found in
Appendices~\ref{sec:cplusplusmodel}, \ref{sec:simpymodel} and
\ref{sec:ciwmodel}.
For the purpose of this paper, Microsoft Excel was chosen as the spreadsheet
software.
Screenshots of the spreadsheet, SIMUL8, and AnyLogic models are shown in
Figures~\ref{fig:spreadsheetscreenshot}, \ref{fig:simul8screenshot}, and
\ref{fig:anylogicscreenshot} respectively.
All models are archived and can be found at~\cite{palmeretal17}.
Some initial observations are summarised in Table~\ref{tab:comparisons_summary}.

\begin{table}
    \begin{center}
        \begin{tabular}{p{0.12\textwidth}p{0.12\textwidth}p{0.13\textwidth}p{0.18\textwidth}p{0.15\textwidth}p{0.09\textwidth}p{0.13\textwidth}}
            \toprule
                                                    & Version controllable            & Licence                                              & Modifiable                            & GUI                                       & Animation                                 & Support \\
            \midrule
            \textbf{\textit{C++}}                   & \cmark                          & GNU GPL free licence.                                & Bespoke models                        & N/A                                       & N/A                                       & N/A \\[1.5cm]
            \textbf{\textit{Spreadsheet modelling}} & \xmark                          & Depends on software.                                 & Limitations to what can be modelled.  & N/A                                       & \xmark                                    & N/A \\[1.5cm]
            \textbf{\textit{SimPy}}                 & \cmark                          & MIT                                                  & Extensible \& modifiable source code. & Limited GUI available for running models. & \xmark                                    & Online documentation \\[1.5cm]
            \textbf{\textit{Ciw}}                   & \cmark                          & MIT                                                  & Extensible \& modifiable source code. & \xmark                                    & \xmark                                    & Online documentation \\[1.5cm]
            \textbf{\textit{AnyLogic}}              & With professional licence only. & Limited PLE version available, otherwise commercial. & Extend with Java.                     & \cmark                                    & \cmark                                    & Online documentation \& paid training. \\[1.5cm]
            \textbf{\textit{SIMUL8}}                & \xmark                          & Commercial                                           & Extend with Visual Logic.             & \cmark                                    & \cmark                                    & Online documentation \& paid training. \\
            \bottomrule
        \end{tabular}
    \end{center}
    \caption{Summary of the comparisons between six simulation frameworks.}
    \label{tab:comparisons_summary}
\end{table}

The capabilities of a spreadsheet model were not found to align with the
expectations of research best practice.
Seeds cannot be set, thus reproducibility is impossible.
The resulting model has very low interpretability, unless set out as a `black
box' model where parameters are input and results are output.
However, this style of model would hinder model understanding and communication.
This in turn leads to low confidence in results, a well reported phenomenon
\cite{ziemannetal16} with the use of spreadsheets.
In addition, most data had to be handled manually, further impeding
reproducibility.

Building a model with a GUI, such as with SIMUL8 and AnyLogic, may ease the
model development process although care should be taken to follow best
practice.
The model may be more accessible given its visual nature which can aid with
communication, although knowledge of the software is required to read much of
the model parameters as many of these are hidden behind objects and menus (for
example, Figures~\ref{fig:simul8screenshot} and~\ref{fig:anylogicscreenshot} do
not in themselves show basic model parameters such as number of servers, arrival
and service rates).
The binary files which represent the SIMUL8 models and the restrictive
commercial licences on both SIMUL8 and AnyLogic inhibit accessibility and model
sharing, and thus reproducibility.

The bespoke model developed in C++, given its compiled nature and that the
model was not held back by unused features, used the least (by far) computation
time.
The model (a script) is shareable and can be put under version control.
Readability is hindered here as all details, including internal simulation
mechanisms, are shown which may make communications with non-specialists
challenging.

The models developed with SimPy and Ciw are version controllable and shareable,
and thus ideal for replicable results.
Compared to the model developed in C++, much of the simulation mechanics are
hidden from the user and pre-tested.
This aids with model communication, validation, and reduces error.
The authors suggest that the Ciw model is more readable than the SimPy version.
Much of the simulation mechanics are on show with SimPy, which increases its
flexibility but reduces readability, whereas Ciw prioritises readability.

\begin{figure}
    \begin{center}
        \includegraphics[width=0.75\textwidth]{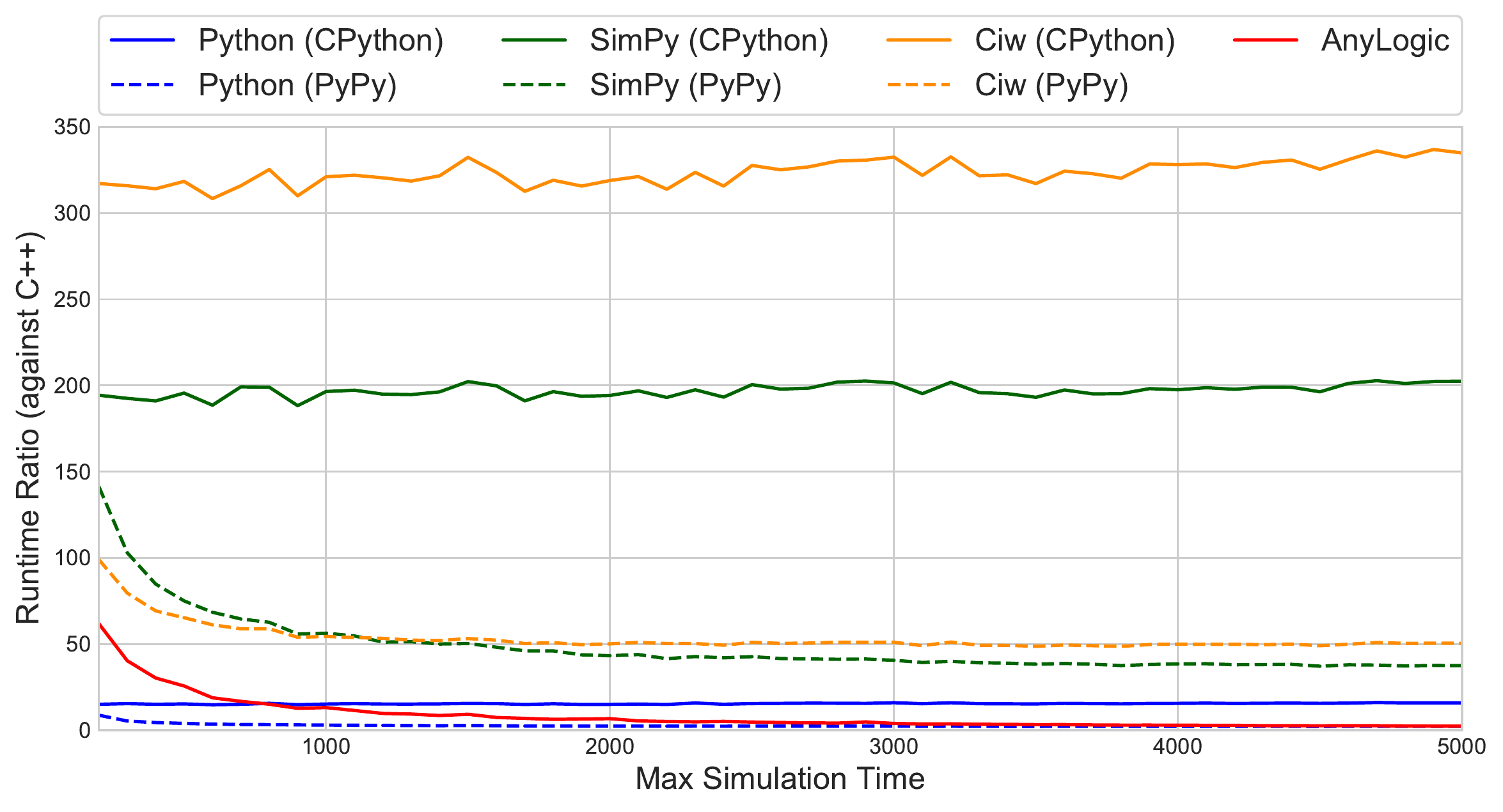}
    \end{center}
    \caption{Comparison between the runtimes of the C++ model, bespoke Python
    model, the SimPy model, the Ciw model, and the AnyLogic model.}
    \label{fig:runtime_analysis}
\end{figure}

Runtime analysis was carried out on the software for which it was possible: C++,
SimPy, Ciw, and AnyLogic.
All analyses were performed on an Apple MacBook Air with 1.4GHz Intel Core i5
processor, OS X 10.9.5 (13F34), with 4 GB 1600 MHz DDR3 memory.
Figure~\ref{fig:runtime_analysis} shows the average runtimes from five runs, as
a ratio of the fastest running model C++, for maximum simulation times from 200
to 5000 time units.
Also included in the analysis is a bespoke Python model (an equivalent model to
the C++ model but coded in Python).
The three Python models were run through two interpreters: the standard CPython
interpreter (version 3.5.1), and PyPy \cite{pypy} (version 5.1.1), which is an
alternative implementation of Python with a Just In Time compiler that improves
runtimes.

It is worth noting that the runtime recorded for the AnyLogic model does not
take into consideration experiment initialisation time, nor does it include
application launch and model loading.
Therefore in practice, running AnyLogic models would take longer than what is
shown in the plot.

Figure~\ref{fig:runtime_analysis} reflects what has been discussed in much of
the literature \cite{law07, robinson14} that bespoke models coded using
programming languages yield faster running models than simulation packages.
The simulation packages here, SimPy and Ciw, carry around many unused features
that may slow down the model.
Running Ciw with PyPy greatly improves performance and we can see that PyPy
seems to be mainly affected by the simulation initialisation time, as running
the simulation for longer sees the performance approaching that of the C++
model.
Furthermore, using Python's inbuilt parallel processing library, it is
straightforward to parallelise trials of Ciw.
Speed, however, is not a priority of Ciw, favoring instead code readability,
and the library's ability to enable reproducible scientific research.
Future development of Ciw may involve performance improvement, as long as this
does not impact readability.
In practice, the computational runtime for Ciw models does not hinder practical
use.

\section{Summary}

This paper has introduced the Python library Ciw for discrete event simulations
of open queueing networks.
The library strives to allow best practices in research software and enable
reproducibility of simulation models.
Ciw has a growing number of features intended to be able to model more complex
systems and ensure that models reflect reality.
It offers advantages over both commercial off the shelf solutions and
programming from scratch.
Table~\ref{tab:pros_cons} summarises the pros and cons of using Ciw.

\begin{table}
\begin{center}
\begin{tabular}{p{0.45\textwidth} p{0.45\textwidth}}
  \toprule
  \textbf{Pros} & \textbf{Cons} \\
  \midrule
  \begin{itemize}
    \item Free (no licence fee, maintenance, or training costs).
    \item Source code open source, thus available for understanding and
    modification.
    \item Full documentation, including tutorials, available online.
    \item Fully and openly tested, giving confidence in use.
    \item Scripting environment offers flexibility in experimentation and
    results analysis.
    \item Can be used in conjunction with other scientific Python tools.
    \item Readable models.
    \item Documented version control enables sustainability.
    \item Continuous development.
    \item Enables testable and version controlled models.
  \end{itemize}
  &
  \begin{itemize}
    \item Some features not available yet.
    \item No GUI or animation currently implemented.
    \item Execution speed compromised for readability.
  \end{itemize} \\
  \bottomrule
\end{tabular}
\end{center}
\caption{Pros and cons of using Ciw.}
\label{tab:pros_cons}
\end{table}

Given that Ciw is built in Python, it is malleable in that it can be combined
with other bespoke functionality with ease.
For example, a user can use inheritance to change the behaviour of a particular
aspect of the system.
This implies that Ciw allows for multi method simulation: for example, it can be
combined with agent based models (inherent to the object-orientated nature of
Python) or combined with some of the ordinary differential equation solvers in
the Python ecosystem (SciPy) to allow for a combination with system dynamics
\cite{robinson14}.

It is hoped others, both academic researchers and practitioners, will make use
of Ciw to conduct reproducible simulations, and also contribute to the library's
development and help add to its growing functionality.

\section{Acknowledgements}

The authors would like to thank Syaribah Brice and Mark Tuson for their
discussions and advice regarding the AnyLogic model.
The authors would also like to thank all contributors to the Ciw project, those
who have written code, and colleagues and users of the library who have
generated ideas and contributed to valuable discussions on the library's
development.
These include: Geraint Palmer, Vincent Knight, Lieke H\"{o}lscher, Sam
Luen-English, Nikoleta Glynatsi, Adam Johnson, Alex Carney, Paul Harper, and
Jennifer Morgan, and a number of users who the authors have only interacted with
online.
A variety of software libraries have been used in this work:

\begin{multicols}{2}
\begin{itemize}
    \item The networkx library (graph theory) \cite{schultetal08}.
    \item The hypothesis library (property based testing) \cite{hypothesis3.8}.
    \item The tqdm library (progress bars) \cite{tqdm}.
    \item The PyYAML library (yaml parsing) \cite{pyyaml}.
    \item The Pandas library (data analysis) \cite{pandas}.
    \item The matplotlib library (data visualisation) \cite{matplotlib}.
\end{itemize}
\end{multicols}

\bibliographystyle{plain}
\bibliography{refs}

\appendix

\section{C++ Model}\label{sec:cplusplusmodel}

\begin{minted}[bgcolor=bg, breaklines=true]{c++}
#include <iostream>
#include <vector>
#include <cmath>
#include <algorithm>
using namespace std;

double runTrial(int seed, double arrivalRate, double serviceRate, int numberOfServers, double maxSimTime, double warmup){
  int i;
  double outcome, r1, r2, serviceTime, serviceStartDate, serviceEndDate, wait;
  double sumWaits = 0.0, arrivalDate = 0.0;
  vector<double> serversEnd;
  vector<double> temp;
  vector<double> waits;
  vector<vector<double> > records;
  srand(seed);

  for(i = 0; i < numberOfServers; ++i){
    serversEnd.push_back(0);
  }

  while(arrivalDate < maxSimTime){
    r1 = ((double)rand() / (double)RAND_MAX);
    r2 = ((double)rand() / (double)RAND_MAX);
    while (r1 == 0.0 || r2 == 0.0 || r1 == 1.0 || r2 == 1.0){
      r1 = ((double)rand() / (double)RAND_MAX);
      r2 = ((double)rand() / (double)RAND_MAX);
    }
    arrivalDate += (-log(r1))/arrivalRate;
    serviceTime = (-log(r2))/serviceRate;
    serviceStartDate = max(arrivalDate, (*min_element(serversEnd.begin(), serversEnd.end())));
    serviceEndDate = serviceStartDate + serviceTime;
    wait = serviceStartDate - arrivalDate;
    serversEnd.push_back(serviceEndDate);
    serversEnd.erase(min_element(serversEnd.begin(), serversEnd.end()));
    temp.push_back(arrivalDate);
    temp.push_back(wait);
    records.push_back(temp);
    temp.clear();
  }

  for(i = 0; i < records.size(); ++i){
    if(records[i][0] > warmup){
      waits.push_back(records[i][1]);
    }
  }

  for(i = 0; i < waits.size(); ++i){
    sumWaits += waits[i];
  }

  outcome = (sumWaits) / (waits.size());
  return outcome;
}

int main(int argc, char **argv){
  int i, seed;
  double solution;
  int numberOfServers = 3;
  int numberOfTrials = 20;
  double arrivalRate = 10.0;
  double serviceRate = 4.0;
  double maxSimTime = 800.0;
  double warmup = 100.0;
  double sumMeanWaits = 0.0;
  vector<double> meanWaits;

  for(seed = 0; seed < numberOfTrials; ++seed ){
    meanWaits.push_back(runTrial(seed, arrivalRate, serviceRate, numberOfServers, maxSimTime, warmup));
  }

  for(i = 0; i < meanWaits.size(); ++i){
    sumMeanWaits += meanWaits[i];
  }
  solution = (sumMeanWaits) / (meanWaits.size());
}
\end{minted}

\section{SimPy Model}\label{sec:simpymodel}

\begin{minted}[bgcolor=bg]{python}
import simpy
import random

arrival_rate = 10.0
number_of_servers = 3
service_rate = 4.0
max_simulation_time = 800
warmup = 100
num_trials = 20

def source(env, arrival_rate, service_rate, server, records):
    """Source generates customers randomly"""
    while True:
        c = customer(env, server, service_rate, records)
        env.process(c)
        t = random.expovariate(arrival_rate)
        yield env.timeout(t)

def customer(env, server, service_rate, records):
    """Customer arrives, is served and leaves."""
    arrive = env.now
    with server.request() as req:
        results = yield req
        wait = env.now - arrive
        records.append((env.now, wait))
        tib = random.expovariate(service_rate)
        yield env.timeout(tib)

def run_trial(seed, arrival_rate, service_rate, number_of_servers,
              max_simulation_time, warmup):
    """Run one trial of the simulation, returning the average waiting time"""
    random.seed(seed)
    records = []
    env = simpy.Environment()
    server = simpy.Resource(env, capacity=number_of_servers)
    env.process(source(env, arrival_rate, service_rate, server, records))
    env.run(until=max_simulation_time)
    waiting_times = [r[1] for r in records if r[0] > warmup]
    return sum(waiting_times) / len(waiting_times)

mean_waits = [run_trial(
  s, arrival_rate, service_rate, number_of_servers,
  max_simulation_time, warmup) for s in range(num_trials)]

average_waits = sum(mean_waits) / len(mean_waits)
\end{minted}

\section{Ciw Model}\label{sec:ciwmodel}

\begin{minted}[bgcolor=bg]{python}
import ciw

max_simulation_time = 800
warmup = 100
num_trials = 20

N = ciw.create_network(
    Arrival_distributions=[['Exponential', 10.0]],
    Service_distributions=[['Exponential', 4.0]],
    Number_of_servers=[3]
)

def run_trial(s, max_simulation_time, warmup):
    """Run one trial of the simulation, returning the average waiting time"""
    ciw.seed(s)
    Q = ciw.Simulation(N)
    Q.simulate_until_max_time(max_simulation_time)
    recs = Q.get_all_records()
    waits = [r.waiting_time for r in recs if r.arrival_date > warmup]
    return sum(waits) / len(waits)

mean_waits = [run_trial(s, max_simulation_time, warmup) for s in range(num_trials)]

average_waits = sum(mean_waits) / len(mean_waits)
\end{minted}

\end{document}